\documentclass[prd,showpacs,nofootinbib]{revtex4}

\usepackage{amsmath}
\usepackage{amsfonts}
\usepackage{amssymb}
\usepackage{bm}

\begin{document}

\title{ Parity-odd and CPT-even electrodynamics of the SME at Finite
Temperature}
\author{Rodolfo Casana, Manoel M. Ferreira Jr, and Madson R. O. Silva}
\affiliation{Departamento de F\'{\i}sica, Universidade Federal do Maranh\~{a}o (UFMA),
Campus Universit\'{a}rio do Bacanga, S\~{a}o Lu\'{\i}s - MA, 65085-580,
Brasil}

\begin{abstract}
This work examines the finite temperature properties of the CPT-even and
parity-odd electrodynamics of the standard model extension. We start from
the partition function written into the functional integral formalism in
Ref. \cite{Finite}. After specializing the Lorentz-violating tensor $%
W_{\alpha \nu \rho \varphi }$ for the nonbirefringent and parity-odd
coefficients, the partition function is explicitly carry out,
showing that it is a power of the Maxwell's partition function.
Also, it is observed that the LIV coefficients induce an anisotropy
in the black body angular energy density distribution. The Planck's
radiation law retains its usual frequency dependence and the
Stefan-Boltzmann law keeps the same form, except for a global
proportionality constant.
\end{abstract}

\pacs{11.30.Cp, 12.60.-i,44.40.+a}
\maketitle

\section{Introduction}

Nowadays, the Standard Model Extension (SME) \cite{string,Kostelecky} is the
theoretical framework most used to investigate Lorentz invariance violation
(LIV). The gauge sector of the SME is composed of a CPT-odd and a CPT-even
sector. The CPT-odd one is constituted by the well-known
Carroll-Field-Jackiw term \cite{CFJ}, whose properties have been well
examined in literature \cite{CFJ2,Consistency,Radiative,Cerenkov}. The
CPT-even part is represent by a tensor $W^{\alpha \nu \rho \varphi }$ which
presents the same symmetries of the Riemann tensor $\left[ W_{\alpha \nu
\rho \varphi }=-W_{\nu \alpha \rho \varphi },W_{\alpha \nu \rho \varphi
}=-W_{\alpha \nu \varphi \rho },W_{\alpha \nu \rho \varphi }=W_{\rho \varphi
\alpha \nu }\right] $ and a double null trace, possessing only 19
independent components.

Recently, the CPT-even has received much attention, yielding the
investigation of new electromagnetic phenomena induced by Lorentz violation
and the imposition of tight upper bounds on the magnitude of the LIV
coefficients. The examination of CPT-even electrodynamics of the SME has
started with Kostelecky and Mewes \cite{KM1} in connection with the study of
polarization deviations for light traveling over large cosmological
distances \cite{KM1,KM2}. Here, one should also mention other researches
involving electromagnetostatics and classical solutions \cite%
{Bailey,Electro,PRD2,Paulo}, radiation spectrum of the electromagnetic field
and CMB \cite{Casana2,Winder,Finite}, photon interactions and quantum
electrodynamics processes \cite%
{Interactions,Inter2,Inter3,Adam2,Klink2,Klink3}, and synchrotron radiation
\cite{Radiation}. A detailed review on the gauge sector of the SME is found
in Ref. \cite{Kostelec}.

At a very recent work \cite{Finite}, we have analyzed the finite temperature
behavior of the parity-even part of the CPT-even sector of the SME as an
attempt to determine the thermodynamics properties of this electrodynamics.
The focus was on the LIV modifications implied on the density energy angular
distribution, the Planck radiation law and implications. The partition
function was written into the functional integral formalism of Matsubara and
explicitly carried out. It was then shown that the altered partition
function is a power of the usual Maxwell's partition function. We have then
observed that, despite small local fluctuations induced by LIV, the Planck
law maintains its usual frequency dependence while the Stefan-Boltzmann
retains its usual $T^{4}$ behavior.

The aim of this present work is to complete the finite temperature
analysis for the CPT-even sector, addressing now the contributions
of the parity-odd components of the tensor $W_{\alpha \nu \rho
\varphi }$ on the thermodynamics of the Maxwell field, searching the
modified Planck's law distribution, angular energy density
distribution, and Stefan-Boltzmann's laws. We thus follow the same
procedure of Ref. \cite{Finite}, taking as starting point the
general partition function attained there. Before being explicitly
evaluated, this partition function shall be specialized for the case
of the nonbirefringent parity-odd coefficients. After explicit
evaluation, we show that the modified partition function is a power
of the Maxwell usual one, in the very same way as observed for the
parity-even case \cite{Finite}.

\section{The theoretical model and results}

The CPT-even gauge Lagrangian of the SME is
\begin{equation}
\mathcal{L}=-\frac{1}{4}F_{\alpha \nu }F^{\alpha \nu }-\frac{1}{4}W^{\alpha
\nu \rho \varphi }F_{\alpha \nu }F_{\rho \varphi },  \label{cpt-1}
\end{equation}%
where $W^{\alpha \nu \rho \varphi }$ is a renormalizable, dimensionless
coupling, composed of 19 elements. In Ref.\cite{Finite}, the Hamiltonian and
constraint structure of this electrodynamics was developed using the Dirac
method. This analysis allowed to write the correct partition function (in
the Matsubara formalism), which integrated on the canonical conjugate
momenta and fields has lead to
\begin{equation}
Z\left( \beta \right) =\det \left( -\square \right) ~\left[ \det \left(
-\square \delta _{ab}+S_{ab}\right) \right] ^{-1/2}.  \label{cpt-33}
\end{equation}%
Here, we define the Euclidean operator, $\square =\partial _{a}\partial
_{a}=\left( \partial _{\tau }\right) ^{2}+\nabla ^{2},$ and the symmetric
Lorentz-violating operator%
\begin{equation*}
S_{ab}=2W_{acdb}\partial _{c}\partial _{d}.
\end{equation*}

Now, we should particularize the tensor $W_{acdb}$\ for the parity-odd
sector, which possesses only three nonbirefringent components. This result
can be achieved using the parametrization of the tensor $W_{\mu \nu \alpha
\beta }$ in terms of four $3\times 3$ matrices,\textbf{\ }$\kappa
_{DE},\kappa _{HB},$ $\kappa _{DB},\kappa _{HE},$ presented in Refs. \cite%
{KM1,KM2}:

\begin{equation}
\left( \kappa _{DE}\right) ^{jk}=-2W^{0j0k},\left( \kappa _{HB}\right) ^{jk}=%
\frac{1}{2}\epsilon ^{jpq}\epsilon ^{klm}W^{pqlm},\left( \kappa _{DB}\right)
^{jk}=-\left( \kappa _{HE}\right) ^{kj}=\epsilon ^{kpq}W^{0jpq}.  \label{P1}
\end{equation}

The matrices $\kappa _{DE}$ and $\kappa _{HB}$ represent the parity-even
sector and possess together 11 independent components, while $\kappa _{DB}$
and $\kappa _{HE}$ stand for the parity-odd described by 8 components. These
four matrices have together the 19 independent elements of the tensor $%
W_{acdb}$. To isolated the parity-odd sector, we take $\kappa _{DE}=\kappa
_{HB}=0.$ The parity-odd sector is written in terms of an antisymmetric ($%
\kappa _{o+})$ and a symmetric matrix $\left( \widetilde{\kappa }%
_{o-}\right) ,$ given as
\begin{equation}
\left( \widetilde{\kappa }_{o+}\right) _{kj}=\frac{1}{2}(\kappa _{DB}+\kappa
_{HE})_{kj},\text{~\ }\left( \widetilde{\kappa }_{o-}\right) _{kj}=\frac{1}{2%
}(\kappa _{DB}-\kappa _{HE})_{kj}.
\end{equation}%
Taking into account the birefringence constraint $\widetilde{\kappa }_{o-}=$
$\frac{1}{2}(\kappa _{DB}-\kappa _{HE})\leq 10^{-32}$\cite{KM1,KM2,Kob}, we
obtain $\kappa _{DB}=\kappa _{HE}.$ This together the condition $\kappa
_{DB}=-\left( \kappa _{HE}\right) ^{T}$ \ implies that the matrix $\kappa
_{DB}=\widetilde{\kappa }_{o+}$ is anti-symmetric (possessing only three
components). Such restriction yields only 3 parity-odd nonbirefringent
parameters, parameterized in terms of a three-vector $\boldsymbol{\kappa }$
\cite{Kob}
\begin{equation}
\kappa _{j}=\frac{1}{2}\epsilon _{jmn}\left( \widetilde{\kappa }_{o+}\right)
_{mn}.
\end{equation}

Into the finite temperature formalism, the matrices (\ref{P1}) are redefined
as
\begin{equation}
\left( \kappa _{DE}\right) _{kj}=2W_{\tau k\tau j},~\ \left( \kappa
_{HB}\right) _{kj}=\frac{1}{2}\epsilon _{kpq}\epsilon _{jmn}W_{pqmn},~\
\left( \kappa _{DB}\right) _{kj}=-\left( \kappa _{HE}\right) _{jk}=W_{\tau
kpq}\epsilon _{jpq}.
\end{equation}

We should now carry out the determinant of the operator $\left( -\square
\delta _{ab}+S_{ab}\right) $ in (\ref{cpt-33}) for the three nonbirefringent
parity-odd components of the tensor $W_{acdb}$, now written in terms of the $%
\boldsymbol{\kappa }$ \ vector as
\begin{equation}
W_{\tau imn}=\frac{1}{2}\left[ \kappa _{m}\delta _{in}-\kappa _{n}\delta
_{im}\right] .  \label{Wodd}
\end{equation}

For computing such functional determinant, we write this operator (in
Fourier space) as $p^{2}\delta _{ab}-\tilde{S}_{ab}$, where $\tilde{S}%
_{ab}=2W_{acdb}p_{c}p_{d}$. Under the prescription (\ref{Wodd}), the matrix
elements of $\tilde{S}_{ab}$ are
\begin{equation}
\tilde{S}_{\tau \tau }=0,~\ \tilde{S}_{\tau j}=\left( \boldsymbol{\kappa }%
\cdot \mathbf{p}\right) p_{j}-\mathbf{p}^{2}\kappa _{j},~\ \tilde{S}%
_{ij}=-2\left( \boldsymbol{\kappa }\cdot \mathbf{p}\right) p_{\tau }\delta
_{ij}+p_{\tau }\left( \kappa _{i}p_{j}+\kappa _{j}p_{i}\right) .
\end{equation}

Thus, the functional determinant is
\begin{equation}
\det \left( -\square \delta _{ab}+S_{ab}\right) =\det \left( -\square
\right) ^{2}\det \left[ -\square -2\left( \boldsymbol{\kappa }\cdot \mathbf{%
\nabla }\right) \partial _{\tau }+\boldsymbol{\kappa }^{2}\mathbf{\nabla }%
^{2}-\left( \boldsymbol{\kappa }\cdot \mathbf{\nabla }\right) ^{2}\right]
\det \left[ -\square -2\left( \boldsymbol{\kappa }\cdot \mathbf{\nabla }%
\right) \partial _{\tau }\right] .
\end{equation}%
Replacing it in the partition function (\ref{cpt-33}), it follows:%
\begin{equation}
Z\left( \beta \right) =Z_{\kappa }^{\left( 1\right) }\left( \beta \right)
Z_{\kappa }^{\left( 2\right) }\left( \beta \right) ,  \label{cpt-37}
\end{equation}%
where the quantities, $Z_{\kappa }^{\left( 1\right) }\left( \beta \right) $
and $Z_{\kappa }^{\left( 2\right) }\left( \beta \right) $, are given as%
\begin{eqnarray}
Z_{\kappa }^{\left( 1\right) }\left( \beta \right) &=&\det \left[ -\square
-2\left( \boldsymbol{\kappa }\cdot \mathbf{\nabla }\right) \partial _{\tau }+%
\boldsymbol{\kappa }^{2}\mathbf{\nabla }^{2}-\left( \boldsymbol{\kappa }%
\cdot \mathbf{\nabla }\right) ^{2}\right]^{-1/2} , \\
Z_{\kappa }^{\left( 2\right) }\left( \beta \right) &=&\det \left[ -\square
-2\left( \boldsymbol{\kappa }\cdot \mathbf{\nabla }\right) \partial _{\tau }%
\right]^{-1/2} ,
\end{eqnarray}
They represent the contributions of the two polarization modes of the
modified photon field. Let us observe that if we only consider the first
order contribution of the LIV background, $\boldsymbol{\kappa }$, both modes
would give the same contribution to the partition function. At leading
order, the associated dispersion relations provide nonbirefringence, a
result in according with the statements of Refs. \cite{KM1,KM2,Kostelec} and
other works that follow this prescription \cite{PRD2,Kob}. The explicit
evaluation of the dispersion relations is developed in the Appendix.

The computation of the functional determinants is performed using the
well-known formulae $\det $\^{O}$=\exp (\mbox{Tr}\ln $\^{O}$)$, thus, we
obtain
\begin{eqnarray}
\ln Z_{\kappa }^{\left( 1\right) }\left( \beta \right)  &=&-\frac{1}{2}\text{%
Tr}\ln \left[ -\square -2\left( \boldsymbol{\kappa }\cdot \mathbf{\nabla }%
\right) \partial _{\tau }+\boldsymbol{\kappa }^{2}\mathbf{\nabla }%
^{2}-\left( \boldsymbol{\kappa }\cdot \mathbf{\nabla }\right) ^{2}\right] ,
\label{Z1k} \\
\ln Z_{\kappa }^{\left( 2\right) }\left( \beta \right)  &=&-\frac{1}{2}\text{%
Tr}\ln \left[ -\square -2\left( \boldsymbol{\kappa }\cdot \mathbf{\nabla }%
\right) \partial _{\tau }\right] .  \label{Z2k}
\end{eqnarray}

We can now evaluate the involved trace of expressions (\ref{Z1k},\ref{Z2k})
writing the gauge field in terms of a Fourier expansion,
\begin{equation}
A_{a}(\tau ,\mathbf{x})=\left( \frac{\beta }{V}\right) ^{\frac{1}{2}}\sum_{n,%
\mathbf{p}}e^{i(\omega _{n}\tau +\mathbf{x}\cdotp\mathbf{p})}\tilde{A}_{a}(n,%
\mathbf{p}),  \label{Fourier1}
\end{equation}%
where $V$ designates the system volume and $\omega _{n}$ are the bosonic
Matsubara's frequencies, $\omega _{n}=\displaystyle\frac{2n\pi }{\beta }$,
for $n=0,1,2,\cdots $.

In this way, the contributions of the two modes of the gauge field
are expressed as
\begin{eqnarray}
\ln Z_{\kappa }^{\left( 1\right) }\left( \beta \right) &=&-\frac{1}{2}V\int
\frac{d^{3}\mathbf{p}}{(2\pi )^{3}}\sum_{m=-\infty }^{+\infty }\ln \beta ^{2}%
\left[ \left( \omega _{m}\right) ^{2}+\mathbf{p}^{2}+2\left( \boldsymbol{%
\kappa }\cdot \mathbf{p}\right) \omega _{m}-\boldsymbol{\kappa }^{2}\mathbf{p%
}^{2}+\left( \boldsymbol{\kappa }\cdot \mathbf{p}\right) ^{2}\right] , \\
\ln Z_{\kappa }^{\left( 2\right) }\left( \beta \right) &=&-\frac{1}{2}V\int
\frac{d^{3}\mathbf{p}}{(2\pi )^{3}}\sum_{m=-\infty }^{+\infty }\ln \beta ^{2}%
\left[ \left( \omega _{m}\right) ^{2}+\mathbf{p}^{2}+2\left( \boldsymbol{%
\kappa }\cdot \mathbf{p}\right) \omega _{m}\right] .
\end{eqnarray}%
For evaluating the integrals, we first implement the translation $\mathbf{p}%
\rightarrow \mathbf{p}-\omega _{m}\boldsymbol{\kappa }$. We then use
spherical coordinates, $\mathbf{p}=\omega \left( \sin \theta \cos \phi ,\sin
\theta \sin \phi ,\cos \theta \right) ,\boldsymbol{\kappa }\cdot \mathbf{p}%
=\kappa \omega \cos \theta ,$ $\omega =\left\vert \mathbf{p}\right\vert ,$ $%
\kappa =\left\vert \boldsymbol{\kappa }\right\vert $. By performing the
summation in $n$, and doing the respective rescalings in the variable $%
\omega $, we obtain the following expressions:
\begin{eqnarray}
\ln Z_{\kappa }^{\left( 1\right) } &=&-\frac{V}{(2\pi )^{3}}\left( 1-\kappa
^{2}\right) ^{3/2}\int d\Omega \frac{1}{\left( 1-\kappa ^{2}\sin ^{2}\theta
\right) ^{3/2}}\int_{0}^{\infty }d\omega ~\omega ^{2}\ln \left( 1-e^{-\beta
\omega }\right) , \\
\ln Z_{\kappa }^{\left( 2\right) } &=&-\frac{V}{(2\pi )^{3}}\left( 1-\kappa
^{2}\right) ^{3/2}\int d\Omega ~\int_{0}^{\infty }d\omega ~\omega ^{2}\ln
\left( 1-e^{-\beta \omega }\right) ,
\end{eqnarray}%
where $d\Omega =\sin \theta d\theta d\phi $ is the solid-angle element.

Then, the partition function for the parity-odd sector of the CPT-even
electrodynamics of the SME is
\begin{equation}
\ln Z=-\frac{V}{(2\pi )^{3}}\left( 1-\kappa ^{2}\right) ^{3/2}\int d\Omega %
\left[ 1+\frac{1}{\left( 1-\kappa ^{2}\sin ^{2}\theta \right) ^{3/2}}\right]
\int_{0}^{\infty }d\omega ~\omega ^{2}\ln \left( 1-e^{-\beta \omega }\right)
.  \label{Zfull-0}
\end{equation}%
The dependence on $\theta $\ shows that the LIV interaction yields an
anisotropic character for the angular distribution of the energy density. By
performing the $\omega -$integration in (\ref{Zfull-0}), we achieve the
energy density per solid-angle element,
\begin{equation}
u\left( \beta ,\Omega \right) \,=\frac{\pi }{120\beta ^{4}}\left( 1-\kappa
^{2}\right) ^{3/2}\left[ 1+\frac{1}{\left( 1-\kappa ^{2}\sin ^{2}\theta
\right) ^{3/2}}\right] ,\   \label{planck2}
\end{equation}%
which reveals the anisotropy induced by the LIV coefficient (the power
angular spectrum is maximal in the plane perpendicular to background
direction). At leading order, the anisotropy factor is quadratic in the $%
\boldsymbol{\kappa }-$vector,
\begin{equation}
u\left( \beta ,\Omega \right) \,\approx \frac{\pi }{120\beta ^{4}}%
\allowbreak \left[ 2+\kappa ^{2}\left( \frac{3}{2}\sin ^{2}\theta -3\right) %
\right] .
\end{equation}%
This result should be contrasted with the linear contribution induced by
anisotropic contribution stemming from the parity-even sector \cite{Finite}.

By performing the angular integrations in Eq. (\ref{Zfull-0}), we find that
the partition function can be written as
\begin{equation}
Z=\left( Z_{A}\right) ^{\gamma (\kappa )},  \label{Zfull-1}
\end{equation}%
where $Z_{A}$ is the partition function of the Maxwell's electrodynamics,
\begin{equation}
\ln Z_{A}=-\frac{V}{\pi ^{2}}\int_{0}^{\infty }d\omega ~\omega ^{2}\ln
\left( 1-e^{-\beta \omega }\right) =V\frac{\pi ^{2}}{45\beta ^{3}}.
\label{ZA}
\end{equation}
and the exponent ${\gamma (\kappa )}$ is a pure function of the LIV
parameter
\begin{equation}
{\gamma (\kappa )}=\left( 1-\kappa ^{2}\right) ^{1/2}\left( 1-\frac{1}{2}%
\kappa ^{2}\right) .  \label{Zfull1a}
\end{equation}

The result (\ref{Zfull-1}) for the nonbirefringent and parity-odd components
of the tensor $W_{\alpha \beta \mu \nu }$ is similar to one obtained in Ref.
\cite{Finite} for the nonbirefringent and parity-even components.

Starting from the equations (\ref{Zfull-0}) or (\ref{Zfull-1}), it is easy
to derive the modified Planck's radiation law or the modifications in the
Stefan-Boltzmann's law, respectively, given as follows:
\begin{equation}
u\left( \omega \right) =\gamma \left( \kappa \right) \frac{1}{\pi ^{2}}\frac{%
\omega ^{3}}{e^{\beta \omega }-1},~\ \ u=\gamma \left( \kappa \right) \frac{%
\pi ^{2}}{15}T^{4}.  \label{planck1}
\end{equation}%
Explicitly, we can observe that the LIV modifications consists in a global
multiplicative function which contain all the LIV correction, this way, the
Planck's radiation law maintains its functional dependence in the frequency
(in all orders in $\kappa )$. Similarly, the energy density\ or the
Stefan-Boltzmann law retains its\ usual temperature dependence ($u\propto {%
T^{4}})$ whereas the Stefan-Boltzmann constant is globally altered as $%
\sigma \rightarrow \gamma \left( \kappa \right) \sigma $, with $\gamma
\left( \kappa \right) $ given by Eq. (\ref{Zfull1a}).

\section{Conclusions and remarks}

In this work we have concluded the study of the finite temperature
behavior of the CPT-even and LIV electrodynamics of the SME which
was started in Ref. \cite{Finite}. We have specialized our analysis
for the nonbirefringent components of the parity-odd sector of the
tensor $W_{\alpha \nu \rho \varphi}$. We have exactly computed the
partition function, (\ref{Zfull-1}), showing that it is a power of
the partition function of the Maxwell electrodynamics as well, being
the power a pure function the LIV parameters. Consequently, the
Planck's radiation law retains its known functional dependence in
the frequency whereas the Stefan-Boltzmann's law keeps the
$T^{4}$-behavior, apart from a multiplicative global factor. It was
observed that the LIV interaction induces an anisotropic angular
distribution for the black body energy density. A similar behavior
was obtained for the nonbirefringent anisotropic components of the
parity-even sector \cite{Finite}. These results show that the
partition function of the full CPT-even sector is expressed as a
power of the Maxwell's one. This pattern, however, is not shared by
the CPT-odd partition function evaluated in Ref.\cite{Casana2}. This
difference is ascribed to the dimensional character of the LIV
coefficient $k_{AF}$.

\appendix

\section{Dispersion relations}

In this Appendix, we write the dispersion relations for this CPT-even and
parity-odd electrodynamics as a procedure to confirm the evaluation of the
associated partition function. It is important to point out that the
dispersion relations of the parity-odd case may be read off directly from
the arguments of the partition functions (\ref{Z1k},\ref{Z2k}) making use of
the prescription $\square \rightarrow -p^{2},\mathbf{\nabla \rightarrow -}i%
\mathbf{p,}$ $\partial _{\tau }\rightarrow -ip_{0},$ which yields
\begin{align}
\left[ p^{2}+2p_{0}\left( \boldsymbol{\kappa }\cdot \mathbf{p}\right) \right]
& =0,  \label{DR1} \\[0.2cm]
\left[ p^{2}+2p_{0}\left( \boldsymbol{\kappa }\cdot \mathbf{p}\right) -%
\boldsymbol{\kappa }^{2}\mathbf{p}^{2}+\left( \boldsymbol{\kappa }\cdot
\mathbf{p}\right) ^{2}\right] & =0.  \label{DR2}
\end{align}

These dispersion relations can be also obtained straightforwardly from the
Maxwell equations for this sector (see Ref. \cite{PRD2}):
\begin{eqnarray}
\nabla \mathbf{\cdot E} &=&-\mathbf{\kappa \cdot }\left( \nabla \mathbf{%
\times B}\right) , \\
\nabla \times \mathbf{B}-\partial _{t}\left( \mathbf{B\times \kappa }\right)
&=&\partial _{t}\mathbf{E}-\nabla \times \left( \mathbf{E\times \kappa }%
\right) \mathbf{,} \\
\nabla \mathbf{\cdot B} &\mathbf{=}&0~,~ \\
\nabla \times \mathbf{E} &=&-\partial _{t}\mathbf{B.}
\end{eqnarray}%
Writing the electric and magnetic fields in a Fourier representation, $%
\mathbf{B}(\mathbf{r})=\left( 2\pi \right) ^{-3}\int \widetilde{\mathbf{B}}(%
\mathbf{p})\exp (-i\mathbf{p}\cdot \mathbf{r})d^{3}\mathbf{p}$, $\mathbf{E}(%
\mathbf{r})=\left( 2\pi \right) ^{-3}\int \widetilde{\mathbf{E}}(\mathbf{p}%
)\exp (-i\mathbf{p}\cdot \mathbf{r})d^{3}\mathbf{p}$, the Maxwell equations
take on the following form (at the absence of sources):
\begin{align}
\mathbf{p}\cdot \widetilde{\mathbf{E}}& =-\mathbf{\boldsymbol{\kappa }}\cdot
\left( \mathbf{p}\times \widetilde{\mathbf{B}}\right) ,  \label{M1} \\[0.2cm]
\mathbf{p}\times \widetilde{\mathbf{B}}+p_{0}\left( \widetilde{\mathbf{B}}%
\times \mathbf{\boldsymbol{\kappa }}\right) +p_{0}\widetilde{\mathbf{E}}& =-%
\mathbf{p}\times \left( \widetilde{\mathbf{E}}\times \mathbf{\boldsymbol{%
\kappa }}\right) \mathbf{,}~  \label{M2} \\[0.2cm]
\mathbf{p}\times \widetilde{\mathbf{E}}-p_{0}\widetilde{\mathbf{B}}& =0,%
\text{ }\mathbf{p}\cdot \widetilde{\mathbf{B}}=0.  \label{M3}
\end{align}

From these expressions, it is attained an equation for the electric field
components, $M^{jl}\widetilde{E}^{l}=0,$ where%
\begin{equation}
M^{jl}=[p^{l}p^{j}-p_{0}p^{j}\kappa ^{l}-p_{0}p^{l}\kappa ^{j}+\delta
^{lj}(p^{2}+2p_{0}A)].  \label{Mij}
\end{equation}%
where $A=\boldsymbol{\kappa }\cdot \mathbf{p}$. Such operator can be
represented as a $3\times 3$ matrix,
\begin{equation}
M^{jl}=\left[
\begin{array}{ccccccc}
p^{2}+2p_{0}A+p_{1}^{2}-2p_{0}p_{1}\kappa _{1} &  &  & p_{1}p_{2}-p_{0}p_{1}%
\kappa _{2}-p_{0}p_{2}\kappa _{1} &  &  & p_{1}p_{3}-p_{0}p_{1}\kappa
_{3}-p_{0}p_{3}\kappa _{1} \\[0.25cm]
p_{1}p_{2}-p_{0}p_{1}\kappa _{2}-p_{0}p_{2}\kappa _{1} &  &  &
p^{2}+2p_{0}A+p_{2}^{2}-2p_{0}p_{2}\kappa _{2} &  &  & p_{2}p_{3}-p_{0}p_{2}%
\kappa _{3}-p_{0}p_{3}\kappa _{2} \\[0.25cm]
p_{1}p_{3}-p_{0}p_{1}\kappa _{3}-p_{0}p_{3}\kappa _{1} &  &  &
p_{2}p_{3}-p_{0}p_{2}\kappa _{3}-p_{0}p_{3}\kappa _{2} &  &  &
p^{2}+2p_{0}A+p_{3}^{2}+2p_{0}p_{3}\kappa _{3}%
\end{array}%
\right] .
\end{equation}%
After suitable simplification, the determinant of this matrix takes the form%
\begin{equation}
\det M^{jl}=p_{0}^{2}\left( p^{2}+2Ap_{0}\right) \left( p^{2}+2Ap_{0}-%
\mathbf{p}^{2}\boldsymbol{\kappa }^{2}+A^{2}\right) .
\end{equation}

The condition $\det M^{jl}=0\ $provides the non-trivial solutions for Eq. (%
\ref{M3}) and the associated dispersion relations of this model,
attained without any approximation. This alternative procedure
confirms the correctness of dispersion relations (\ref{DR1},
\ref{DR2}) and of the expressions (\ref{Z1k}, \ref{Z2k}), written at
the finite temperature formalism. The intricate character of the
relations (\ref{DR1}, \ref{DR2}), involving both $p_{0}$ and
$\mathbf{p,}$ imply the following dispersion
relations:%
\begin{eqnarray}
\omega _{\pm } &=&-\left( \boldsymbol{\kappa }\cdot \mathbf{p}\right) \pm
\sqrt{\mathbf{p}^{2}+\left( \boldsymbol{\kappa }\cdot \mathbf{p}\right) ^{2}}%
,  \label{E01} \\
\omega _{\pm } &=&-\left( \boldsymbol{\kappa }\cdot \mathbf{p}\right) \pm
\sqrt{\mathbf{p}^{2}(1+\boldsymbol{\kappa }^{2})},  \label{E02}
\end{eqnarray}%
which are different even at leading order in $\mathbf{\kappa .}$ Assuming $|%
\mathbf{\kappa |<<}1,$ the expressions (\ref{E01},\ref{E02}) are reduced to
the form:
\begin{align}
\omega _{+}& =|\mathbf{p|}-\left( \boldsymbol{\kappa }\cdot \mathbf{p}%
\right) ,  \label{E1A} \\[0.08in]
\omega _{-}& =|\mathbf{p|}+\left( \boldsymbol{\kappa }\cdot \mathbf{p}%
\right) .  \label{E1B}
\end{align}%
Here, the root ~$\omega _{+}=|\mathbf{p|}-\left( \boldsymbol{\kappa }\cdot
\mathbf{p}\right) $ represents a positive frequency mode, since~$|\mathbf{%
\kappa |<<}1.$ On the other hand, the mode $\omega _{-}=(|\mathbf{p|}+\left(
\boldsymbol{\kappa }\cdot \mathbf{p}\right) )$ stands for the positive
energy of an anti-particle (after reinterpretation). This is a negative
frequency mode. It should be mentioned that, despite the double sign in the
dispersion relations (\ref{E1A},\ref{E1B}), they yield the same phase
velocities for waves traveling at the same direction. Note the the positive
and negative frequency modes are associated with waves which propagate in
opposite directions and the term $\left( \boldsymbol{\kappa }\cdot \mathbf{p}%
\right) $ changes of the signal under the direction inversion ($\mathbf{%
p\rightarrow -p).}$ This result confirms the nonbirefringent character of
the coefficient $\boldsymbol{\kappa } $ at leading order, as properly stated
in Refs. \cite{KM1,KM2,Kostelec}, and others \cite{PRD2,Kob}. These same
dispersion relations can obtained by means of a general evaluation for the
dispersion relations (see Appendix of Ref. \cite{Finite}) or by means of the
poles of the gauge propagator of this electrodynamics (see Ref. \cite%
{Propagator}).

\begin{acknowledgments}
R. C. thanks Conselho Nacional de Desenvolvimento Cient\'{\i}fico e Tecnol%
\'{o}gico (CNPq) for partial support, M. M. F. is grateful CNPq and to
FAPEMA (Funda\c{c}\~{a}o de Amparo \`{a} Pesquisa do Estado do Maranh\~{a}o).
\end{acknowledgments}

\end{document}